\documentclass[article,prl,twocolumn,superscriptaddress,showpacs,nofootinbib,showkeys,10pt]{revtex4}
\usepackage{graphicx}  
\usepackage{amsmath}   
\usepackage{amssymb}   
\usepackage{bm} 
\usepackage{dcolumn}
\usepackage{color}
\usepackage{mathrsfs}
\usepackage{amsfonts}
\usepackage{varioref}
\usepackage{textcomp}
\RequirePackage[colorlinks,citecolor=blue,urlcolor=magenta,linkcolor=blue]{hyperref}
\allowdisplaybreaks

\labelformat{section}{Section #1} 
\labelformat{subsection}{Section #1} 
\labelformat{subsubsection}{Section #1}
\labelformat{subsubsubsection}{Section #1}
\labelformat{equation}{Eq.~(#1)} 
\labelformat{figure}{Fig.~#1} 
\labelformat{subfigure}{Fig.~\thefigure#1} 
\labelformat{table}{Tab.~#1} 
\labelformat{appendix}{Appendix #1}
\begin{document}
\title{Stimulated Hawking Emission From Electromagnetic Analogue Black Hole:\\ Theory and Observation }
\author{Avijit Bera}
\email{avijitbera1997@gmail.com}
\affiliation{Physics and Applied Mathematics Unit, Indian Statistical Institute, Kolkata-700108, India}
\author{Subir Ghosh}
\email{subirghosh20@gmail.com}
\affiliation{Physics and Applied Mathematics Unit, Indian Statistical Institute, Kolkata-700108, India}

\begin{abstract}
In this letter we  consider possible analogue Hawking radiation from a normal dielectric and metamaterial composite, having an  analogue horizon where the dielectric parameters vanish and change sign upon crossing this transition zone. We follow a complex path analysis to show the presence of an analogue Hawking temperature at the horizon and subsequent photon production from the ambient electromagnetic field. Possibility of experimental observation is also commented upon.
\end{abstract}
\maketitle
{\bf Introduction :}  In this letter we show that an electromagnetic Black Hole (BH) constructed out of graded metamaterial is capable of undergoing stimulated Hawking emission. Specifically we predict the Hawking temperature of this analogue BH to be 
\begin{equation}\label{30}
	T = \frac{c \hbar}{4\pi k r_0} ,
\end{equation}
where $c$ and $k$ are velocity of light in vacuum and  Boltzmann constant respectively. $r_0$ is the radial position of the effective horizon, which is the transition zone between ordinary dielectric and metamaterial. The dielectric parameters change from positive (normal dielectric) to negative (metamaterial) values across the horizon. Of particular interest is the fact that precisely such a system was theoretically proposed by Narimanov and Kildishev \cite{kil} and experimentally realized in laboratory by Cheng, Cui, Jiang and  Cai \cite{lab}  few years back (for review see \cite{rev lab}). Both the theory \cite{kil} (see also \cite{more}) and observation \cite{lab} of  BH nature of the system, {\it i.e.} omnidirectional absorption of light with nearly 100-percent efficiency, is entirely classical involving Maxwell equations in an inhomogeneous medium. In this work we extend the analysis to quantum (or semi-classical, to be specific) regime and the possibility of stimulated Hawking radiation of photons emerges. Indeed, this is one more example of {\it  analogue} Hawking effect.
The quantum analysis exploits the complex path formalism of Landau and Lifshitz \cite{lan}, as adapted by Srinivasan and  Padmanabhan \cite{pad} in the context of Hawking radiation.

Even after around 45 years, Hawking's theoretical discovery \cite{haw} of BH radiation remains in the limelight in a somewhat  enigmatic way since no smoking gun evidence of it has been produced in the astrophysical context, the major reason being the tiny value of Hawking temperature $\sim 10^{-8}K $ for a solar mass black hole. This has led to a flurry of activity  in the analogue gravity scenario, pioneered by Unruh \cite{un} in fluid system, followed by many works in different physical systems (see  \cite{rev} for a review). In all these cases the underlying principle is that the relevant degree of freedom of a physical system, (that need not have any direct connection to General Relativity or Gravitation and preferably possible to construct the system in a laboratory), satisfies an equation of motion that can be cast in  the conventional form of dynamics in an effective curved spacetime. This allows the identification of an effective metric constructed out of parameters of the analogue system. However, an analogue black hole and subsequently analogue Hawking Radiation (HR) is possible only if the effective metric has a ``horizon'' where the metric coefficients vanish or become singular, as the case may be. The ``dumb hole'' theorized by Unruh \cite{un} and later experimentally constructed and studied by \cite{oth} occurred in a moving fluid (with a critical velocity) that acoustic disturbance can not penetrate. (In fact the experimental setups are more geared to an analogue white hole).

Let us come to optical analogue of BH, that is of present interest. It has long been well known \cite{lan} (in purely classical framework) that electrodynamics in curved spacetime can be mapped to electrodynamics in flat spacetime in a dielectric medium with non-uniform permittivity  and permeability. This theme received a strong impetus in recent times after the laboratory construction of an exotic form of dielectric - {\it metamaterial}, with negative permittivity  and permeability. Possible existence  of metamaterial was visualized long ago by Veselago \cite{ves} and its experimental verification was done in \cite{expt}. Its counterintuitive properties have led to many significant applications such as
negative refraction,
electromagnetic invisibility cloaks, super-resolution imaging, electromagnetic concentrators
and light trapping, among others \cite{rev appl}. The essential features of the constitutive parameters can be induced in 
periodic or non-periodic arrays of electric or magnetic resonant or non-resonant particles.
Electrodynamics in medium with both normal and metamaterial features has become a popular playground for theorists since exploiting the above curved spacetime-dielectric mapping the latter can serve as analogue models of exotic forms of spacetime \cite{smol}. 

Specific theoretical suggestion and possible construction of an omnidirectional electromagnetic absorber in the microwave frequency that can serve as an optical analogue BH appeared in \cite{kil}. Laboratory construction \cite{lab, rev lab} of a graded metamaterial disc with rotational symmetry has demonstrated BH-like features by absorbing incident radiation extr emely efficiently. The system, with a metamaterial core inside, can trap and absorb electromagnetic waves coming from all directions that spiral inwards without any reflections. The theoretical study (and the subsequent experiment), is  based on a purely classical analysis of the Maxwell equations in a non-uniform dielectric medium. We emphasize that this is a optical analogue of a classical BH and so analogue Haking effect, a quantum phenomenon, has not been considered. An  attempt of metamaterial BH with possible Hawking effect is given in \cite{gm}. In an earlier work \cite{rez} a similar idea was put forward without explicitly using the metamaterial framework. In a dynamical system, using perturbative quantum field theory framework, analogue Hawking effect was predicted by \cite{oth1} and experimentally observed by \cite{other} where  non-linear Kerr effect was induced in the medium by a laser pulse.

In this letter, in a completely novel analysis, we show that the model of optical analogue BH proposed and constructed  in \cite{kil,lab,rev lab}  can generate stimulated HR. The difference between the two is that in BH  HR \cite{haw}  is emitted spontaneously at the cost of lowering the BH mass whereas in stimulated HR the source of analogue HR is the ambient electromagnetic field. The dielectric-metamaterial interface creates an effective horizon that allows emergence of photons only whereas in BH HR all forms of particles are allowed to emerge. We have gone beyond the classical analysis \cite{kil} to a semi-classical analysis where quantum tunnelings are involved. Precisely the latter are taken care of by introducing the complex path formalism \cite{lan1}.

The original work of Hawking used Bogoliubov coefficients but later identical results have been recovered in different frameworks such as tunneling \cite{wil} picture, anomaly \cite{wil1}, Hamilton-Jacobi (HJ) with complex path \cite{pad}. This clearly underlines the universality of HR phenomenon. We will use the HJ approach initiated by Srinivasan and Padmanabhan \cite{pad} because of its generality. Generally in HR (analogue HR) one deals with the field equation in curved metric (effective metric) respectively that has horizon (effective horizon). We, on the other hand, will start directly from electromagnetic wave equation, derived from Maxwell's equations, and show that a horizon-like property of the dielectric parameters can induce an analogue Hawking temperature and analogue HR since the solution of the HJ equation has an imaginary part developed in the transition zone.

The paper is organized as follows:  discussion of electrodynamics in (spherically symmetric) non-uniform dielectric medium;  Hamilton-Jacobi formalism for the specific form of normal-metamaterial composite; brief discussion on analogue Schrodinger problem. Then we provide the main result:  derivation of  analogue Hawking temperature for the graded metamaterial composite and  its possible experimental observation. The letter ends with a discussion and possible future directions of research.
\vskip .5cm
{\bf Electrodynamics in non-uniform medium :}
Maxwell's equations in dielectric medium free of charge and current density are given by
\begin{equation}\label{01} 
\nabla . \textbf{D} = 0 ,~
\nabla . \textbf{B} = 0 \nonumber $$$$
\nabla \times \textbf{E} = -\frac{\partial \textbf{B}}{\partial t},~ 
\nabla \times \textbf{H} = -\frac{\partial \textbf{D}}{\partial t}. 
\end{equation}
In spherical polar coordinates $r,\theta,\phi$ let us define the displacement vector $\textbf{D}$ and magnetic induction $\textbf{B}$ from the electric field $\textbf{E}$ and magnetic field $\textbf{H}$ as
\begin{equation}
\label{1}
D_r=\epsilon _1 E_r, \: 
D_\theta =\epsilon _2 E_\theta, \: 
D_\phi =\epsilon _2 E_\phi
\end{equation}
\begin{equation}
\label{2}
B_r=\mu _1 H_r, \:
B_\theta =\mu _2 H_\theta, \:
B_\phi =\mu _2 H_\phi 	
\end{equation}	
where we restrict to $\epsilon_i(r), \mu_i(r)$, ie.  permittivity $\epsilon$ and permeability $\mu$ are both the functions of radial distance $r$ only. This yields 
\begin{equation}\label{3}
\nabla .\textbf{E} = -\frac{E_r}{\epsilon_1} \partial_r \epsilon_1 + (1-a)\frac{1}{rsin\theta} 	\partial_\theta(sin\theta E_\theta)  $$$$ + (1-b)\frac{1}{rsin\theta}\partial_\phi E_\phi	
\end{equation}	
where $a=\frac{\epsilon_2}{\epsilon_1}$ and $b=\frac{\epsilon_3}{\epsilon_1}$. In a similar way we obtain, with $p=\frac{\mu_2}{\mu_1}, q=\frac{\mu_3}{\mu_1}$, 
\begin{equation}\label{4}
\nabla .\textbf{H} = -\frac{H_r}{\mu_1} \partial_r \mu_1 + (1-p)\frac{1}{rsin\theta} 	\partial_\theta(sin\theta H_\theta) $$$$+ (1-q)\frac{1}{rsin\theta}\partial_\phi H_\phi .	
\end{equation}	
For simplicity we restrict further to $a=b=1$ and also $p=q=1$ leading to 
\begin{equation}\label{5}
\nabla . \textbf{E}	= -\frac{1}{\epsilon(r)} \nabla \epsilon(r).\textbf{E}~~, ~~	\nabla .\textbf{H} = -\frac{1}{\mu(r)} \nabla \mu(r).\textbf{H} .
\end{equation}
It is straightforward to recover the wave equation satisfied by $\textbf{E}$ as (with a similar equation for $\textbf{H}$),
\begin{equation}\label{6}
\nabla ^2 \textbf{E} -\mu \epsilon \frac{\partial^2 \textbf{E}}{\partial t^2}	+ \nabla (\frac{1}{\epsilon}) (\nabla \epsilon.\textbf{E}) + (\frac{1}{\epsilon}) \nabla (\nabla \epsilon.\textbf{E})=0.
\end{equation}
Exploiting spherical symmetry of our system, we  expand the electric field in terms of vector spherical harmonics:
\begin{equation}\label{7}
\textbf{E} = \sum_{l=0} ^ \infty \sum_{m=0} ^ \infty [E^r _{lm} (r) \textbf{Y}_{lm} + E^1 _{lm} (r) \Psi_{lm} + E^2 _{lm} (r) \Phi_{lm} ]	
\end{equation}
where $E^r_{lm}(r)$ is the radial component and $E^1 _{lm} (r)$ , $E^2 _{lm} (r)$ are the transverse components of the electric field. The vector spherical harmonics are defined in terms of conventional spherical harmonics as 
\begin{eqnarray}
\textbf{Y}_{lm} &=& Y_{lm} \hat{r} \nonumber \\
\Psi_{lm} &=& r \nabla Y_{lm} = [\hat{\theta}\frac{\partial}{\partial \theta} + \hat{\phi} \frac{1}{sin \theta} \frac{\partial}{\partial \phi}] Y_{lm} \nonumber \\
\Phi_{lm} &=& \textbf{r} \times \nabla Y_{lm} = [\hat{\phi}\frac{\partial}{\partial \theta} - \hat{\theta} \frac{1}{sin \theta} \frac{\partial}{\partial \phi}] Y_{lm} \nonumber~.
\end{eqnarray}
After a fair amount of messy algebra  we isolate wave equations of the individual components of $\textbf{E}$, 
\begin{equation}\label{08}
\frac{\partial^2 E^r_{lm}}{\partial r^2} - \mu(r)\epsilon(r) \frac{\partial^2 E^r_{lm}}{\partial t^2} + \{  \frac{2}{r} + \frac{1}{\epsilon(r)} \frac{\partial \epsilon(r)}{\partial r}\} \frac{\partial E^r_{lm}}{\partial r}  $$$$ + \frac{l(l+1)}{r^2}(2E^1_{lm} - E^r_{lm}) + \{ \frac{\partial}{\partial r} (\frac{1}{\epsilon(r)} \frac{\partial \epsilon(r)}{\partial r}) - \frac{2}{r^2}\} E^r_{lm} = 0
\end{equation}	

\begin{equation}\label{09}
[\frac{1}{r^2}\{ \frac{\partial}{\partial r} (r^2 \frac{\partial E^1_{lm}}{\partial r}) - l(l+1)E^1_{lm}  \} -\frac{E^1_{lm}}{r^2 sin^2\theta} $$$$ -\mu(r)\epsilon(r) \frac{\partial^2 E^1_{lm}}{\partial t^2}   
+\{\frac{2}{r^2}+ \frac{1}{r \epsilon(r)}\frac{\partial \epsilon(r)}{\partial r}\} E^r_{lm}] $$$$ \times  \{ sin\theta (\frac{\partial Y_{lm}}{\partial \theta})^2 + \frac{1}{sin \theta} (\frac{\partial Y_{lm}}{\partial \phi})^2\}  $$$$
-\frac{2cos\theta}{r^2 sin^2\theta}[E^1_{lm}\frac{\partial Y_{lm}}{\partial \theta} + \frac{E^2_{lm}}{sin\theta} \frac{\partial Y_{lm}}{\partial \phi} ] = 0	
\end{equation}	
There is a similar equation for $E^2_{lm}$. Let us concentrate on the $s$-wave mode that is $l=m=0$ for which (Eq. 9) identically vanish and the $E^r_{00}\equiv E$-equation (Eq. 8) is considerably simplified,
\begin{equation}\label{11}
E'' - \mu\epsilon \frac{\partial^2 E}{\partial t^2} + \{  \frac{2}{r} + \frac{\epsilon '}{\epsilon}  \} E'  +  \{(\frac{\epsilon '}{ \epsilon})' - \frac{2}{r^2}\} E = 0 ,
\end{equation}
where $()'\equiv \partial_r()$. 

As an aside we mention an intriguing result: how will the wave behave in a medium whose dielectric parameters are obtained from the Schwarzschild BH metric
\begin{equation}\label{24}
dS^2 = g_{\mu \nu} dx^\mu dx^\nu = (1-\frac{R}{r})dt^2 + \frac{dr^2}{(1-\frac{R}{r})} $$$$ + r^2(d\theta ^2 + sin^2\theta d\phi ^2 )
\end{equation}
through the mapping \cite{lan}
\begin{equation}\label{25}
\epsilon^{ij} = \mu^{ij} = - \frac{\sqrt{-g}}{g_{00}} g^{ij}.
\end{equation}
Since the mapping is in cartesian coordinate system 	the corresponding parameters are explicitly written as 
\begin{equation}\label{26}
\epsilon^{ij}|_{Cart.} = \mu^{ij}|_{Cart.} = - \frac{1}{1-\frac{R}{r}} (\delta^{ij} -\frac{x^i x^j}{r^3} R) , ~~ x^i = x,y,z.
\end{equation}
Returning to polar coordinates we find (details are in Supplemental material B)
\begin{equation}\label{new}
\epsilon^{ij}|_{polar} = \mu^{ij}|_{polar} =diag~\epsilon_0 [1,~B^{-1}(r),~B^{-1}(r)].
\end{equation}
The subsequent wave equations are (full expressions are in Supplemental material)
\begin{eqnarray}\label{40}
\epsilon_0 ^2 \partial_t ^2 E_r - \nabla ^2 E_r = [F]
\end{eqnarray}
\begin{equation}\label{41}
\frac{\epsilon_0 ^2}{B} \partial_t ^2 E_\theta - \nabla ^2 E_\theta = [G],~~
\frac{\epsilon_0 ^2}{B} \partial_t ^2 E_\phi - \nabla ^2 E_\phi = [H]
\end{equation}
where $F,G,H$ comprise of components or electric vector and their {\it spatial} derivatives. $E_\theta , E_\phi $ are linear combinations of our previous parameterization $E^1,E^2$. Notice that only the all important time derivative part of $E_\theta , E_\phi $ in (\ref{41}) contains the horizon factor $B(r)$ and hence will participate in the quantum tunneling whereas $E_r$ in (\ref{40})  will not contribute to tunneling which is somewhat counterintuitive. 
\vskip .5cm
{\bf Graded metamaterial  and Hamilton-Jacobi scheme :}
Let us consider $\epsilon,\mu$ to be time-independent real functions  of the form
\begin{equation}\label{12}
\epsilon(r)=\frac{\epsilon_0}{B(r)}~,~~\mu(r)=\frac{\mu_0}{B(r)}
\end{equation}
where $\epsilon_0,\mu_0$ denote constants (values at vacuum) with $c=1/\sqrt{(\epsilon_0\mu_0)}$ denoting the velocity of electromagnetic wave in vacuum and  $B(r)$ having a "horizon" at $r_0$ satisfying $B(r_0)=0$ with $B'(r_0)$ being non-vanishing and finite.

The semiclassical or WKB wave-function follows from making the standard ansatz 
\begin{equation}\label{13}
E(r,t)=exp[\frac{i}{\hbar}S(r,t)].
\end{equation}
Thus   (\ref{11}) leads to the dynamical equation  for $S$,
\begin{equation}\label{14}
\frac{1}{(cB)^2}(\frac{i}{\hbar}\ddot S -\frac{1}{\hbar ^2}(\dot S)^2)-
(\frac{i}{\hbar}S'' -\frac{1}{\hbar ^2}(S')^2)-\frac{i}{\hbar}\rho(r)S' $$$$ -\sigma(r)=0
\end{equation}
where $\rho =(\frac{2}{r}-\frac{B'}{B}),~\sigma =[(\frac{B'}{B})^2-\frac{B''}{B}-\frac{2}{r^2}]$ and $\dot {()}\equiv \partial_t()$.
As is customary \cite{pad}, the expansion of $S$ in powers of $\hbar/i$,
\begin{equation}\label{15}
S(r,t)=S_0(r,t)+(\frac{\hbar}{i})S_1(r,t)+(\frac{\hbar}{i})^2S_2(r,t)+~....
\end{equation}
is substituted in (\ref{14}) thereby yielding the lowest order in $\hbar$  equation,
\begin{equation}\label{16}
\frac{E^2}{(cB)^2}-(S_0')^2 + \hbar^2\sigma(r)=0.
\end{equation}
In the above $E$ is a constant, the particle energy, obtained from the standard HJ identification $\dot S_0=-E$. Notice that, although of $O(\hbar^2)$, we have kept $\hbar^2\sigma(r)$ because it involves $1/B(r)$ which is singular at the horizon. The solution of the HJ equation or equivalently the semiclassical wave-function is derived as 
\begin{equation}\label{17}
S_0(r,t)=-Et\pm \frac{1}{c}\int \frac{dr}{B}{\sqrt {E^2-\hbar^2c^2B^2\sigma}}~.
\end{equation}
Since $B^2\sigma$ is finite at the horizon we can safely ignore $\hbar^2B^2\sigma$ because in calculating Hawking effect the neighborhood of horizon is all that matters. Thus $S_0$ simplifies to
\begin{equation}\label{18}
S_0(r,t)=-Et\pm \frac{E}{c}\int \frac{dr}{B}~.
\end{equation}
Before proceeding further, as is advocated in \cite{pad}, we check the validity of the semiclassical ansatz by computing $S_1$ obtained from the next-to-leading order (in $\hbar$) equation from (\ref{14}),  
\begin{equation}\label{19}
2(\frac{1}{(cB)^2}\dot S_0 \dot S_1 - S_0'  S_1')+ \frac{1}{(cB)^2}\ddot S_0 - S_0'' -\rho S_0' =0.
\end{equation}
From (\ref{18}) we obtain 
\begin{equation}\label{20}
\dot S_0=-E,~\ddot S_0=0,~S_0'=\pm \frac{E}{cB},~S_0''=\mp \frac{EB'}{cB^2}= \mp \frac{ER}{cB^2},
\end{equation}
where near horizon $B(r)\approx B'(r_0)(r-r_0)\equiv R(r-r_0)$. From (\ref{19}) we compute the $O(\hbar)$ correction to $S_0$, with $\dot S_1=-E_1t$,
\begin{equation}\label{21}
S_1(r,t)=-E_1t+\int dr(\pm \frac{E_1}{cB}+\frac{R}{2B}+\frac{\rho}{2}).
\end{equation}
Since all the terms in the integrand have same order of singularity at horizon as that of $S_0$, the validity of the semiclassical approximation is assured. We will be calculating the analogue Hawking temperature from (\ref{18}).
\vskip .5cm
{\bf Effective Schrodinger problem :}
It is also instructive to take a detour \cite{pad} to see that the above problem can be cast in to a one-dimensional Schrodinger problem by making the substituting 	
\begin{equation}\label{22}
E(r,t)=exp[\frac{i}{\hbar}Et]\frac{\sqrt B}{r}\Psi(r) ,
\end{equation}
in (\ref{11}). $\Psi $ satisfies a time-independent Schodinger equation, (in the limit $\hbar \rightarrow 0$),
\begin{equation}\label{23}
-\Psi ''-\frac{g}{(r-r_0)^2}\Psi=0
\end{equation}
where $g=(E/(c\hbar R))^2$. This equation is valid near the horizon with $(r-r_0)^{-2}$ dominating over other terms in the potential.  We derive an inverted oscillator pot ential similar to \cite{pad}. To exploit the standard machinery   the trick \cite{pad} is to introduce an energy $\tilde E$ by replacing (\ref{23}) by 
\begin{equation}\label{24}
-\Psi ''-\frac{g}{(r-r_0)^2}\Psi=\tilde E\Psi
\end{equation}
and subsequently take the limit $\tilde E\rightarrow 0$. We will not repeat the steps of \cite{pad} which will yield same result as (\ref{18}) that we will be following.
\vskip .5cm
{\bf Analogue Hawking temperature and  analogue Hawking radiation :}
As we have emphasized throughout beauty and utility of the complex path formalism \cite{pad} is its generality. Operationally, we claim that since the relevant (or analogue) equations are structurally identical to the BH context similar conclusions should prevail. Again we will skip the detailed discussion provided in \cite{pad} and only present a few salient steps. 

We will consider the function $B(r)$ such that it mimics the Schwarzschild-like metric where $B(r)\ge 0$ for $r\ge r_0$ and  $B(r)\le 0$ for $r\le r_0$ with the horizon at $r=r_0,~B(r_0)=0$. Thus in the disc shaped dielectric, the core region $r_0\ge r \ge 0$ will be of metamaterial type with negative dielectric parameters whereas the region $r\ge r_0$ will consist of normal dielectric.  This will allow us to follow the same reasoning (as for Hawking radiation)  where we treat the light going out from metamaterial to normal medium as outgoing and light from normal to metamaterial zone as incoming. The former and latter correspond to positive and negative sign respectively in the integral term of (\ref{18}). Infact the optical black hole laboratory system \cite{lab} is of this form. 

Thus recalling (\ref{18}) for photons  crossing the analogue horizon from the core $r\le r_0$  to $r\ge r_0$ (with $(\partial S_0)/(\partial r)\ge 0$ we have
\begin{equation}\label{25}
S_{out} ^0 = -  \frac{E_1}{c} \int_{r_0-\epsilon}^{r_0+\epsilon} \frac{dr}{B}|_{\epsilon \rightarrow 0} +(real~part)  $$$$
=\frac{i\pi E}{R(r_0)}+(real~part) .
\end{equation}
Similarly,   for photons  crossing the analogue horizon from outside $r\ge r_0$ in to the core $r\le r_0$ (with $(\partial S_0)/(\partial r)\le 0$ we have
\begin{equation}\label{26}
S_{in} ^0 = -  \frac{E_1}{c} \int_{r_0+\epsilon}^{r_0-\epsilon} \frac{dr}{B}|_{\epsilon \rightarrow 0} +(real~part)  $$$$
=-\frac{i\pi E}{R(r_0)}+(real~part) .
\end{equation}
The modulus square of the above expressions yield the respective probabilities,
\begin{equation}\label{27}
P[out]~\infty ~ exp(-\frac{2\pi E}{\hbar R}) ,~~	P[in]~\infty ~ exp(\frac{2\pi E}{\hbar R}),
\end{equation} 
which in tern means that 
\begin{equation}\label{28}
P[out] = \exp (-\frac{4\pi E}{\hbar R})~ P[in].
\end{equation}
This is written as $	P[out] = \exp (-\beta E)~ P[in]$ with the analogue Hawking temperature of the horizon  identified as
\begin{equation}\label{29}
\beta = \frac{1}{kT} = \frac{4\pi}{c \hbar R} 
\end{equation}	
where $k$ is the Boltzmann constant. If we consider $B(r)$ of the Schwarzschild form $B(r)=1-r_0 /r$ then the analogue Hawking temperature at $r=r_0$ becomes
\begin{equation}\label{30}
T = \frac{c \hbar}{4\pi k r_0} .
\end{equation}
The above (\ref{29}) constitute our main result.
Expressing in another way for  $T \approx 1K$ requires $r_0 \sim 10^{-3}~meter $  and $T \approx 300K$ will need the composite disc to have  $r_0 \sim 3\times 10^{-6}~meter $. A more realistic choice  might be $B=b~tanh(\frac{r-r_0}{r_0})$ where $b$ is a dimensionless constant. in this case $R=b/r_0$ yielding thereby
\begin{equation}\label{31}
T=b~\frac{c \hbar}{4\pi k r_0}.
\end{equation}
The possibility of experimental observation of Hawking effect from disc-shaped metamaterial composite appears to be quite promising. The main challenge  seems to be the dimension of the system since the analogue Hawking temperature $T\sim (r_0)^{-1}$, larger $T$ requires a smaller $r_0$, the analogue horizon. In the case of \cite{lab} the disc diameter is $\sim 10^{-3}~meter$s.

\vskip .5cm
{\bf Discussion :}
In this letter we have shown the potential possibility of  Hawking-like junction temperature and subsequent Hawking-like radiation from the junction of a composite normal-metamaterial dielectric. We have used the complex path formalism \cite{lan1}, closely following the work \cite{pad} as applied to Hawking radiation.  We have exploited the generality involved in the formalism that requires only an effective horizon-like structure in the wave equation of the relevant degree of freedom involved. Intuitively one can understand the special status of the analogue horizon and the need to invoke tunneling mechanism to cross the horizon in following ways. On one hand, the effective (phase) velocity of the wave  $\sim B(r)/\sqrt{\epsilon_0\mu_0}$ vanishes at the horizon. On the other hand, in the geometrical optics limit we have the relation $p=\sqrt{(\epsilon(r)/\epsilon_0)}E/c=\sqrt{(1/B(r))}E/c$ where (for the zero angular momentum case we are restricting to), $p$ and $E$ correspond to the photon momentum and energy (or frequency) respectively. Thus, outside the horizon $p$ becomes imaginary and classically forbidden and the photons have to tunnel out. Let us end our work on a cautionary note keeping in mind the recent paper \cite{yar} that critically analyses the possibility of Hawking effect in other analogue systems.

The ambient electromagnetic field energy will act as the source of this radiation. 
It should be noted that, in the present paper, we have not taken into account two very important aspects of metamaterial: it is inherently dispersive and  lossy in nature. To address these issues, the dielectric parameters need to be frequency dependent and complex. Hence our model and results are actually valid for a narrow range of frequency with negligible loss. We plan to return to these topics in near  future.
\vskip .5cm
{\bf Acknowledgments :}	
We thank  S. Shankaranarayanan, Bibhas Ranjan Majhi,  Yaron Kedem, Tie Jun Cui for correspondence and Alexander V. Kildishev for informing us of relevant papers.

\end{document}